\documentstyle[11pt,aaspp4]{article}

\begin{document}

\title{Day-Scale Variability of 3C~279 and Searches for 
Correlations in Gamma-Ray, X-Ray, and Optical Bands}

\author{R.~C.~Hartman\altaffilmark{1,2},
           M.~Villata\altaffilmark{3},
        T.~J.~Balonek\altaffilmark{4},
        D.~L.~Bertsch\altaffilmark{1},
           H.~Bock\altaffilmark{5},
           M.~B\"ottcher\altaffilmark{6,7},
	   M.~T.~Carini\altaffilmark{18},
           W.~Collmar\altaffilmark{8},
           G.~De~Francesco\altaffilmark{3},
        E.~C.~Ferrara\altaffilmark{9},
           J.~Heidt\altaffilmark{5},
           G.~Kanbach\altaffilmark{8},
           S.~Katajainen\altaffilmark{10},
           M.~Koskimies\altaffilmark{10},
        O.~M.~Kurtanidze\altaffilmark{17},
           L.~Lanteri\altaffilmark{3},
           A.~Lawson\altaffilmark{12},
        Y.~C.~Lin\altaffilmark{13},
        A.~P.~Marscher\altaffilmark{14},
        J.~P.~McFarland\altaffilmark{9},
        I.~M.~McHardy\altaffilmark{12},
        H.~R.~Miller\altaffilmark{9},
           M.~Nikolashvili\altaffilmark{11},
           K.~Nilsson\altaffilmark{10},
        J.~C.~Noble\altaffilmark{14},
           G.~Nucciarelli\altaffilmark{15},
           L.~Ostorero\altaffilmark{3},
           T.~Pursimo\altaffilmark{10},
        C.~M.~Raiteri\altaffilmark{3},
           R.~Rekola\altaffilmark{10},
           T.~Savolainen\altaffilmark{10},
           A.~Sillanp\"a\"a\altaffilmark{10},
           A.~Smale\altaffilmark{16},
           G.~Sobrito\altaffilmark{3},
        L.~O.~Takalo\altaffilmark{10},
        D.~J.~Thompson\altaffilmark{1},
           G.~Tosti\altaffilmark{15},
        S.~J.~Wagner\altaffilmark{5},
        J.~W.~Wilson\altaffilmark{9}}

\altaffiltext{1}{Code 661, NASA/GSFC, Greenbelt, MD 20771}
\altaffiltext{2}{rch@egret.gsfc.nasa.gov}
\altaffiltext{3}{Osservatorio Astronomico di Torino, Strada
                           Osservatorio 20, I-10025 Pino Torinese, Italy}
\altaffiltext{4}{Department of Physics and Astronomy, Colgate University, 
                                  13 Oak Drive, Hamilton, NY 13346-1398}
\altaffiltext{5}{Landessternwarte K\"onigstuhl, 69117 Heidelberg, Germany}
\altaffiltext{6}{Department of Space Physics and Astronomy, Rice
				     University, Houston, TX 77005-1892}
\altaffiltext{7}{Chandra Fellow}
\altaffiltext{8}{Max-Planck-Institut f\"ur Extraterrestrische Physik,
                                  P.O. Box 1603, 85740 Garching, Germany}
\altaffiltext{9}{Department of Physics and Astronomy, Georgia State
                                            University Atlanta, GA 30303}
\altaffiltext{10}{Tuorla Observatory, V\"ais\"al\"antie 20,
                                            FIN-21500 Piikki\"o, Finland}
\altaffiltext{11}{Abastumani Observatory, 383762 Abastumani,
						     Republic of Georgia}
\altaffiltext{12}{Department of Physics and Astronomy, University of
                                                         Southampton, UK}
\altaffiltext{13}{W. W. Hansen Experimental Physics Laboratory, Stanford
                                          University, Stanford, CA 94305}
\altaffiltext{14}{Institute for Astrophysical Research, Boston
                     University, 725 Commonwealth Ave., Boston, MA 02215}
\altaffiltext{15}{Osservatorio Astronomico di Perugia, Via Bonfigli,
                                                    06123 Perugia, Italy}
\altaffiltext{16}{Code 662, NASA Goddard Space Flight Center, 
						     Greenbelt, MD 20771}
\altaffiltext{17}{Astrophysikalisches Institut Potsdam, An der Sternwarte
                                              16, 14482 Potsdam, Germany}
\altaffiltext{18}{Dept. of Physics and Astronomy, Western Kentucky
		      University, 1 Big Red Way, Bowling Green, KY 42104}

\begin{abstract}

Light curves of 3C~279
are presented in optical (R-band), X-rays (RXTE/PCA), and
$\gamma$~rays (CGRO/EGRET) for 1999 Jan--Feb and 2000 Jan--Mar.  During
both of those epochs the $\gamma$-ray levels were high, and all three
observed bands demonstrated substantial variation, on time scales as
short as one day.  Correlation analyses provided no consistent pattern,
although a rather significant optical/$\gamma$-ray correlation was seen
in 1999, with a $\gamma$-ray lag of $\sim$2.5~days, and there are other
suggestions of correlations in the light curves.  For comparison, 
correlation analysis is also presented for the $\gamma$-ray and X-ray
light curves during the large $\gamma$~ray flare in 1996 Feb and the two 
$\gamma$-bright weeks leading up to it; the correlation at that time was
strong, with a $\gamma$-ray/X-ray offset of no more than 1 day.

\end{abstract}


\section{Introduction}

The discovery by the EGRET instrument on the Compton Gamma Ray
Observatory (CGRO) that blazars can be strong $\gamma$-ray emitters
posed an intriguing question: what is the mechanism responsible for
this previously unknown and sometimes dominant high-energy emission?
Obvious candidates are the well-known synchrotron-self-Compton 
(SSC), and external-Compton (EC) models; in both of these suggested
processes, the synchrotron-emitting relativistic electrons would
energize soft photons via the inverse-Compton process.  In this
scenario, the principal
matter of debate is the origin of the soft photons, the 
choices being synchrotron photons alone (SSC; e.g. Marscher \& Gear
1985; Maraschi et al. 1992; Bloom \& Marscher 1996), photons of
different provenance, i.e., accretion disk (ECD; Dermer et al. 1992;
Dermer \& Schlickeiser 1993; Sikora et al. 1994) or broad-line
region (ECR; Blandford \& Levinson 1995; Ghisellini \& Madau 1996;
Dermer et al. 1997), or a combination of those possibilities.

Other possibilities, which have not yet been as carefully explored,
are the proton-driven models.  In the proton-initiated cascade (PIC)
scenario (Mannheim 1993), very high-energy
protons initiate photopion production, resulting in a
$\gamma$-ray/electron/positron cascade.  More recently, other
proton-driven models have been discussed (e.g. Protheroe 1996a,
1996b, Rachen 2000, Aharonian 2000, M\"ucke \& Protheroe 2001).  
These seem to be more readily applied to the lower-luminosity
blazars (often described as high-frequency-peaked blazars, HBL's), 
but might also be adaptable for 
higher-luminosity blazars such as 3C~279.

Since the models predict different relationships between variations
in the different observing bands, intensive simultaneous monitoring
in several widely-spaced bands can provide crucial information on
the radiation mechanisms and the structure of the jet.  For that
reason, several coordinated multiwavelength campaigns have been
carried out in the last years on EGRET-detected blazars.

We present here the results of simultaneous monitoring in three
bands, GeV $\gamma$~rays (from the CGRO/EGRET instrument),
X-rays (from the PCA instrument
on the RXTE satellite), and R-band optical (from a number of
observers and ground-based observatories).  For comparison, we
also show a similar analysis of the $\gamma$-ray and X-ray
light curves from the three weeks leading up to and including
the large $\gamma$-ray flare in early February of 1996 (Wehrle
et al. 1998).  There was little optical coverage at that time, but
the $\gamma$~rays and X-rays were strongly correlated.

\section{Observations}

The campaigns were centered around the following observation 
sequences:

1996 Jan 16 -- Feb 06 (CGRO viewing periods 511.0, 511.5)

1999 Jan 20 -- Feb 01 (CGRO viewing periods 806.5, 806.7)

2000 Feb 09 -- Mar 01 (CGRO viewing periods 910.0, 911.1)

CGRO viewing period 806.5 was an observation of 3C~273 during which
EGRET was scheduled to be off.  Because of the optical brightness of
3C~279, EGRET was turned on in full-field-of-view-mode to see if 
3C~279 was detectable in the EGRET energy range.  It was indeed bright,
which led to the implementation of a target-of-opportunity observation,
viewing period 806.7, with 3C~279 better centered in the field of view.

CGRO viewing period 910.0 was a 3C~279 target of opportunity, based on
optical activity and brightness.  Viewing period 911.1 was an
observation of 3C~273 during which EGRET was scheduled to be off;
due to the high $\gamma$-ray level seen in vp 910.0, EGRET was left
on for vp 911.1, and was switched to full-field-mode to maximize
the sensitivity to 3C~279, which was well off-axis.

\subsection{Optical}

R-band observations were made at a number of observatories, as 
described below, during both the 1999 and the 2000 campaigns.  The 
coverage thus provided was the best 
ever obtained on a flat-spectrum radio quasar (FSRQ) during an EGRET
observation.  Most of the observations were made from European
observatories.  Several of the authors and observatories are members
of the WEBT consortium.

3C~279 was observed during 1999 Jan 18 -- Feb 13 and 
2000 Jan 29 -- Feb 19 at the Abastumani Astrophysical Observatory
(Republic of Georgia) using a Peltier-cooled ST-6 CCD camera
attached to the Newtonian focus of the 70~cm meniscus telescope (1/3).
The full frame field of view is 14.9$\times$10.7~arcmin$^2$.
All observations are performed using combined filters of glasses which
match the standard B, V (Johnson) and $R_C, I_C$ (Cousins) bands well.
Because the scale of the CCD and the meniscus telescope resolution are
2.3$\times$2.7~arcsec$^2$ per pixel and 1.5~arcsec respectively, the images
are undersampled; therefore the frames were slightly defocused to satisfy
the sampling theorem.  A full description of the Abastumani blazar
monitoring program is given in Kurtanidze \& Nikolashvili (1999).

Observations in 2000 were made with the 60~cm KVA telescope on La Palma,
Canary Islands, using a ST-8 CCD camera with BVR filters.  The data 
reduction was done using IRAF (with bias and flatfield corrections).

Observations were taken with the 1.2~m telescope of Calar Alto 
Observatory, Spain and with the 0.7~m telescope of the 
Landessternwarte Heidelberg.  Both telescopes are equipped with 
$LN_2$-cooled CCD cameras.  Observations in Heidelberg are carried 
out with a Johnson R band filter.  The Calar Alto observations were
carried out in Johnson R (in 2000) and R\"oser R (in earlier years). 
Standard de-biasing and flat-fielding was carried out before 
performing differential aperture photometry.  (Finding charts and
comparison sequences are available at
http://www.lsw.uni-heidelberg.de/projects/extragalactic/charts.html
for 3C~279, along with many other blazars.)

Observations were performed using Lowell Observatory's 42~inch
Hall telescope and the 24~inch telescope of the Mount Stromlo / 
Siding Spring Observatories.  Both telescopes are equipped with a 
direct CCD camera and an autoguider. The observations were made
through VRI filters.  Repeated exposures of 90~s were obtained
for the star field containing 3C~279 and several comparison stars 
(Smith et al. 1985).  These comparison
stars were internally calibrated and are located on the same CCD
frame as 3C~279.  They were used as the reference standard stars in 
the data reduction process.  The observations were reduced following
Noble et al. (1997), using the method of Howell and Jacoby
(1986).  Each exposure is
processed through an aperture photometry routine which reduces the 
data as if it were produced by a multi-star photometer. Differential
magnitudes can then be computed for any pair of stars on the frame.
Thus, simultaneous observations of 3C~279, several comparison stars,
and the sky background will allow one to remove variations which may be
due to fluctuations in either atmospheric transparency or extinction.
The aperture photometry routine used for these observations is the
{\it phot} task in IRAF.

Observations were taken with the 2.5~m Nordic Optical Telescope (NOT) 
on La Palma, Canary Islands, Spain, using the ALFOSC instrument with 
a 2000$\times$2000 CCD camera (0.189~arcsec per pixel), and V and R-filters.
Data reduction (including bias and flat field corrections) were made
either with standard IRAF or MIDAS (J. Heidt) routines.

Observations at the Perugia Observatory were carried out with the
Automatic Imaging Telescope (AIT).  The AIT is based on an 
equatorially mounted 40~cm f/5 Newtonian reflector.  A CCD camera 
and Johnson-Cousins $BVR_cI_c$ filters are
utilized for photometry (Tosti et al. 1996).  The data were reduced
using aperture photometry with the procedure described in that
reference.

Observations at the Torino Observatory were done with the 1.05~m REOSC 
telescope. The equipment includes an EEV CCD camera (1296$\times$1152 
pixels, 0.467~arcsec per pixel) and standard (Johnson-Cousins) $BVRI$ 
filters. Frames are reduced by the Robin procedure locally developed 
(Lanteri 1999), which includes bias subtraction, flat fielding, and 
circular Gaussian fit after background subtraction.  The magnitude 
calibration was performed according to the photometric sequence by 
Raiteri et al. (1998). Magnitudes were converted to fluxes by using a
B-band Galactic extinction of 0.06 mag and following Rieke \& Lebofsky 
(1985) and Cardelli et al. (1989).

\subsection{X-Rays}

3C~279 was the target for a series of 36 RXTE monitoring observations
during 1999 January 2 -- February 16, for a total on-source time
of 67~ks.  The X-ray data presented here were obtained using the
Proportional Counter Array (PCA) instrument in the Standard 2 and Good
Xenon configurations, with time resolutions of 16~s and $<1\mu$s
respectively.  Only PCUs 0, 1, and 2 were
reliably on throughout the observations, and we limit our analysis to
data from these detectors.

A further sequence of 28 monitoring observations was performed with
RXTE in 2000 February, using the same instrumental configurations, for
a total on-source time of 104~ks.  For this sequence we utilized
data from PCUs 0 and 2.
 
Data analysis was performed using RXTE standard analysis
software, FTOOLS 5.0.  Background subtraction of the PCA data was
performed utilizing the ``L7-240'' models generated by the RXTE
PCA team.  The quality of the background subtraction was checked in
two ways: (i) by comparing the source and background spectra and light
curves at high energies (50--100~keV) where the source itself no longer
contributes detectable events; and (ii) by using the same models to
background-subtract the data obtained during slews to and from the
source.

\subsection{Gamma Rays}

The EGRET instrument is sensitive to $\gamma$~rays in the energy
range 30 to 30,000~MeV.  Its capabilities and calibration are described
in Thompson et al. (1993), Esposito et al. (1999), and Bertsch 
(2001).  Point source data are analyzed using likelihood techniques 
(Mattox et al. 1996).  The choice of one day as the unit of 
integration is dictated by the sensitivity of the EGRET detector
and the general level of the emission during the time intervals
of these observations.  In TeV $\gamma$~rays, significant variations
have been seen on time scales well under one hour (REFS),
and may very well be present in 3C~279 in GeV $\gamma$-rays also.

The 1999 and 2000 $\gamma$-ray data presented here have been shown 
previously in a preliminary form in Hartman et al. 2001a \& 2001b.
Unfortunately, in both of those references, there was a 1-day error in
the Julian Dates for the $\gamma$-ray observations.  Thus the
$\sim$3.5-day optical to $\gamma$-ray lag in 1999 tentatively 
reported there corresponds to the $\sim$2.5-day lag discussed here.

\subsection{Light Curves}

The light curves resulting from the observations described above are
shown in Figures 1 and 2.

\section{Correlation Analysis and Results}

Although the $\gamma$-ray observations were continuous, both the R-band
and X-rays were only sampled, sometimes irregularly and/or sparsely; 
thus the discrete correlation
function (DCF), which was designed for analysis of unevenly sampled
data (Edelson \& Krolik 1988), was used for this analysis.  Because of
the large statistical errors on the EGRET data points, the initial
analysis was done using equation (3) of Edelson \& Krolik (1988); this
resulted in unphysical normalization in the correlation results.
According to J.~Krolik (private communication), this is a known but
unresolved effect in DCF analyses.  Reanalysis ignoring the errors on
the EGRET data points resulted in reasonable normalizations, and
produced equally significant correlations.  Therefore the correlation
results shown below all ignore the EGRET errors.

Figures 3 and 4 show the results of the DCF analyses for the 1999 and
2000 light curves, respectively.  The following is a summary of the
evidence found for correlations.

\subsection{1999}

\subsubsection{$\gamma$-ray/X-ray}

Two possible correlations are found, in which the
$\gamma$~rays lag the X-rays by about 10 and 5-6 days.  The first links 
the highest two X-ray points with the two $\gamma$-ray peaks; the second
links the first $\gamma$-ray peak with the second X-ray high point and
the second (sharp) $\gamma$-ray peak with a time period with little
X-ray coverage.
While mathematically possible, these correlations are unconvincing
because of the very limited X-ray coverage around the relevant times.
In addition, the long delays are probably difficult to account for
theoretically.

\subsubsection{$\gamma$-ray/optical}

If a 2--3 day $\gamma$-ray lag is assumed, the
$\gamma$-ray light curve is very similar to that in the R-band, and
the DCF analysis finds this to be a rather strong correlation.  It
is the most convincing correlation found in the six DCF analyses
for 1999 and 2000.

A negative correlation with an 8--9 day optical lag links the first
$\gamma$-ray peak with the R-band minimum around TJD 210 and the second
$\gamma$-ray peak in the optically unsampled TJD 213--218 interval.
In addition to being unconvincing because of coverage limitations,
this seems quite unphysical.

\subsubsection{X-ray/optical}

A possible correlation with a 2.5-day optical lag
requires that the two optical peaks around TJD 194 and 198 be linked to
the two highest X-ray points, and ignores the strongest and sharpest
optical feature, placing it in the weak X-ray minimum of TJD 202.
This is unconvincing because of the very limited coverage around the
two X-ray high points, and also because the most prominent optical
feature is ignored.

A fairly significant correlation with an 7--8 day optical lag links the 
two highest X-ray points with the optical peaks at about TJD 198.5
and 204.0 .  This is conceivable, but in addition to the poor X-ray
sampling around the relevant times, the 7--8 day offset seems 
difficult to accommodate theoretically.

\subsection{2000}

\subsubsection{$\gamma$-ray/X-ray}

At zero time-delay, there is a sharp peak in the DCF; its statistical
significance is only about 1.8$\sigma$, but the correlated pattern
is obvious to the eye in the light curves, not only around the sharp
peaks, but in the entirety of both light curves.

The variations seen here are more complicated than those seen in both
$\gamma$-rays and X-rays in 1996 Jan--Feb, when a zero time-delay
was also seen.  

\subsubsection{$\gamma$-ray/optical}

No significant correlation was found.

\subsubsection{X-ray/optical}

A possible correlation with a one-day optical lag seems
plausible, but ignores the X-ray peak around TJD 585.

Another possible correlation is seen with an 11-day X-ray lag. 
This is rather unconvincing because of X-ray variations that do not 
show up in the optical, and is also somewhat implausible physically.

\subsection{1996 High State and Large Flare}

For comparison, DCF analysis is presented of the $\gamma$-ray and X-ray
light curves for the three weeks in 1996 Jan--Feb leading up to and
including the large $\gamma$-ray flare.  The light curves, adapted
from Wehrle et al. 1998, are shown in Figure 5.  
To the eye, the $\gamma$~rays and X-rays appear well-correlated, with
no apparent lag;
this is confirmed by the DCF analysis, the results of which are shown
in Figure 6.

\section{Discussion and Conclusions}

In the 1999 and 2000 light curves,
the correlation that is most apparent to the eye is that in 1999 
between the $\gamma$~rays and the optical, with a $\sim$2.5-day
$\gamma$-ray lag.  With the $\gamma$-rays, peaked around TJD 206.5, 
apparently correlating with the optical peak around TJD 204, it is 
not easy to imagine a scenario that could produce such a sequence 
(a discussion on various
possibilities can be found in Hartman et al. 2001a, 2001b).

Another correlation that seems apparent to the eye, that of the
X-rays and $\gamma$~rays in 2000, gives a disappointingly weak
effect ($\sim$2$\sigma$ for zero delay) when analyzed with the DCF,  
as noted above.  Examination of the autocorrelations in those
two bands (Figure 7)
provides some assistance in interpreting the results.  Although
the X-ray autocorrelation is fairly routine, that for the
$\gamma$-rays is unusual.  Apparently this is due to the two
one-day high points separated by a day for which the best $\gamma$-ray
flux estimate is zero, albeit with substantial statistical errors
on all of the points.  (Examination of the photon maps for the
three days under discussion verifies that the $\gamma$-rays do
disappear during the middle day.)

Thus there is no consistent pattern found.  This could be because 
the emission in the three bands investigated really does have no
persistent relationship, or merely because the data are not
adequate, in coverage and/or statistical accuracy, to bring out such
relationships.

What correlations and time delays are to be expected here?
Detailed predictions about the theoretically expected light
curves are difficult because of the multitude of physical
processes potentially involved in the formation and evolution
of the particle and photon spectra, in particular in
FSRQ's.  While detailed modeling of
variability patterns expected in high-frequency peaked
BL Lac objects (which are well modeled with pure SSC
models) has been done (e.g., Takahashi et al. 1996,
Georganopoulos \& Marscher 1998, Kusunose, Takahara \& Li 2000,
Li \& Kusunose 2000), detailed theoretical work relevant to the 
short-term variability of the multi-component spectra probably 
present in the high-energy emission from FSRQ's is still in its
very early stages (see, e.g., Sikora et al. 2001).  Thus we 
must restrict the discussion of the expected time lags to 
order-of-magnitude estimates at this point. 

Frequency-dependent time lags in the short-term variability of 
FSRQ's like 3C~279 are likely to be related to either the electron
cooling in the effectively emitting region, or the dynamical time 
scale on which the soft seed photon fields for Compton scattering
are changing in the frame of the relativistically moving emitting
region.  The time scale for acceleration of relativistic electrons
might be of the order of the time scale for Fermi acceleration,
$\tau_{\rm acc} \sim {2 \pi r_L / c} \sim 3.6 \times 10^{-7}
\gamma B$~s in the co-moving frame of the emitting region
(where $r_L$ is the Larmor radius, $\gamma$ is the electron Lorentz 
factor, and $B$ is the magnetic field in G), or
$\tau^{\ast}_{\rm acc} = {\tau_{\rm acc} / D}$ in the observer's 
frame (where $D \sim $10 is the Doppler boosting factor determining 
the time contraction between the co-moving and the observer's 
frame).  Thus, for any reasonable value of the magnetic field, 
variability on the acceleration time scale will be smeared out by 
light travel time effects, and would be too short to be resolvable 
with current multiwavelength observations anyway.

Based on the multi-epoch multiwavelength spectral fits to 3C~279
presented in Hartman et al. (2001c), we can estimate the typical
electron cooling time scale in the emitting region, assuming that
(as indicated by the spectral fits) electron cooling is dominated
by inverse-Compton scattering of external radiation fields.  Taking
into account both the contributions from direct accretion disk
radiation and from reprocessing of this radiation within the broad
line region, we find the observed cooling time is
$$ \tau_{\rm cool}^{\ast} = \left( {4 \over 3} \, c \, \sigma_T 
{u_s \over m_e c^2} \, \gamma \, D \right)^{-1} \, ,$$
where $\sigma_T$ is the Thomson cross section, and the energy density 
of soft photons in the emitting region is given by
$$ u_s \approx {L_D \over 4 \pi c} \left( {1 \over z^2 \, \Gamma^2} +
{\Gamma^2 \, \tau_{\rm blr} \over r_{\rm blr}^2} \right) \, .$$
Here $L_D$ is the accretion disk luminosity, $z$ is the distance
of the emitting region from the central engine, $\Gamma$
is the bulk Lorentz factor of the emitting region, and
$\tau_{\rm blr}$ and $r_{\rm blr}$ are the radial Thomson depth 
of the broad line region and its average distance
from the central engine, respectively.
For reasonable values of the parameters ($L_D \sim 10^{46}$~erg/s,
$z \sim 0.025$~pc, $\Gamma \sim 10$, $\tau_{\rm blr} \sim 0.003$,
and $r_{\rm blr} \sim 0.1$~pc), we find that the cooling time scale 
relevant to electrons emitting in the EGRET energy regime 
($\gamma \gtrsim 10^4$) is of the order of one to several hours, 
while for particles emitting predominantly at X-ray and optical 
frequencies ($\gamma \lesssim 100$), it is expected to be one
to several days,
which would then be the relevant time scale determining time lags 
between different energy bands. 

If time lags are dominated by the dynamical time scale on which the
soft seed photon fields (for Compton scattering, to produce the 
high-energy radiation) are changing, we would expect 
typical time lags of
$$ \tau_{\rm dyn}^{\ast} \sim {\Delta z (1 - \beta_{\Gamma} \cos\theta )
\over c} \, .$$
For typical values of the parameters ($\Delta z \sim 0.1$~pc,
$D \sim \Gamma \sim 10$, and $\cos\theta \sim \beta_{\Gamma}$, 
this gives delays of a few days.  Thus, in both cases the expected 
time lags between different photon energy bands would be of the 
order of 1 to a few days.

In HBL's, proton-driven models can produce time delays of the order of
a day or less (A. M\"ucke, private communication).  This is probably
true also of the more complicated FSRQ's such as 3C~279,
but so far there has been no theoretical study demonstrating this.

It should be noted that, despite the similar flux levels and variations
in 1999 and 2000, Hartman et al. (2001c) have shown evidence that
conditions in the inner region of 3C~279 may have been substantially
different in 2000 than in 1999.  In particular, the strength of the
broad line emission may have been much weaker in 2000.

Several strong implications for future investigations such as this
are clear:

$\gamma$~rays - Two necessary improvements are obvious:
considerably better statistics and longer observation intervals.
Both of these requirements will be met by the GLAST mission, planned
for launch in 2006:
(1) The GLAST observation plan (at least for the first year or two), is
to operate in a scanning mode, so that a large fraction of the sky will 
be covered during each orbit.  Objects near the equator, such as 3C~279,
will receive good exposure on every orbit during the scanning part of
the mission, and the entire sky will receive significant coverage over
one day; (2) Its larger effective area, larger field of view, and 
better point spread function (compared with EGRET), will provide much 
better statistics and sensitivity than are available from EGRET.

The Italian $\gamma$-ray telescope AGILE (Vercellone et al. 1999), 
planned for launch in
2002-2003, will have sensitivity comparable to that of EGRET, but
with significantly better angular resolution.  Its observing
program will permit longer observations than were usually possible
with EGRET.

X-rays - The obvious need here is for more uniform coverage.
Unfortunately, it is unlikely that the RXTE satellite, with its
very flexible scheduling, will be operating by the time of the GLAST
launch.  The big X-ray missions expected to be operating in parallel
with GLAST are likely to be less flexible and accessible than RXTE.
They will, however, have much greater sensitivity, permitting
investigation of dimmer objects than RXTE.
The X-ray monitors on HETE II and Swift may be able to provide some
assistance, but their availability and applicability are unclear at
the present time.
Thus good X-ray time-sampling during the GLAST era appears uncertain at
present.

Optical - Although the optical coverage was good during most of the time
during and around the EGRET observations used here, there were some
substantial holes that allowed the DCF to suggest unlikely
correlations.  Future investigations of this type will certainly
need to improve upon this.  Some of the intensive optical monitoring
presented here utilized automated telescopes.  Hopefully, additional
automated systems, at sites throughout the world, will be available
by the time of the GLAST launch.

The WEBT (Whole Earth Blazar Telescope; Mattox 1999a, 1999b;
Villata et al. 2000) is a
different approach to intensive optical monitoring.  It is a
consortium of about twenty optical observatories around the world,
formed to facilitate 24-hour high-time-density blazar observations
during multiwavelength campaigns.  For additional information see
http://swampfox.fmarion.edu/~jmattox/webt/ .

\acknowledgments

The work at Torino Observatory and Perugia University Observatory
was partly supported by the Italian
Ministry for University and Research (MURST) under grant Cofin98-02-32 
and by the Italian Space Agency (ASI).

The Nordic Optical Telescope is operated on the island of La Palma 
jointly by Denmark, Finland, Iceland, Norway and Sweden in the 
Spanish Observatorio del Roque de
Los Muchachos of the Instituto de Astrofisica de Canarias.
The Tuorla Observatory authors wish to thank The Finnish Academy for
support.

H. Bock, J. Heidt and S.J. Wagner acknowledge support by the DFG 
(SFB 328 and 439), and CAHA/DSAZ
for support during several observing runs on Calar Alto.

O.M. Kurtanidze thanks the Astrophysikalisches Institute Potsdam for 
support.

The Georgia State University authors wish to thank Lowell and
Mount Stromlo / Siding Spring Observatories for allocations of 
observing time.  This work has been supported in part by an award 
from GSU's RPE Fund to PEGA, and by grants from the Research 
Corporation and NASA (NAGW-4397).

The work of M.~B\"ottcher is supported by NASA through Chandra
Postdoctoral Fellowship Award No. 9-10007, issued by
the Chandra X-ray Center, which is operated by the
Smithsonian Astrophysical Observatory for and on behalf
of NASA under contract NAS 8-39073.

IRAF is distributed by the National
Optical Astronomy Observatories, which is operated by the Association
of Universities for Research in Astronomy, Inc., under cooperative
agreement with the National Science Foundation.

\clearpage
\begin{figure}
\figurenum{1}
\epsscale{1.5}
\plotfiddle{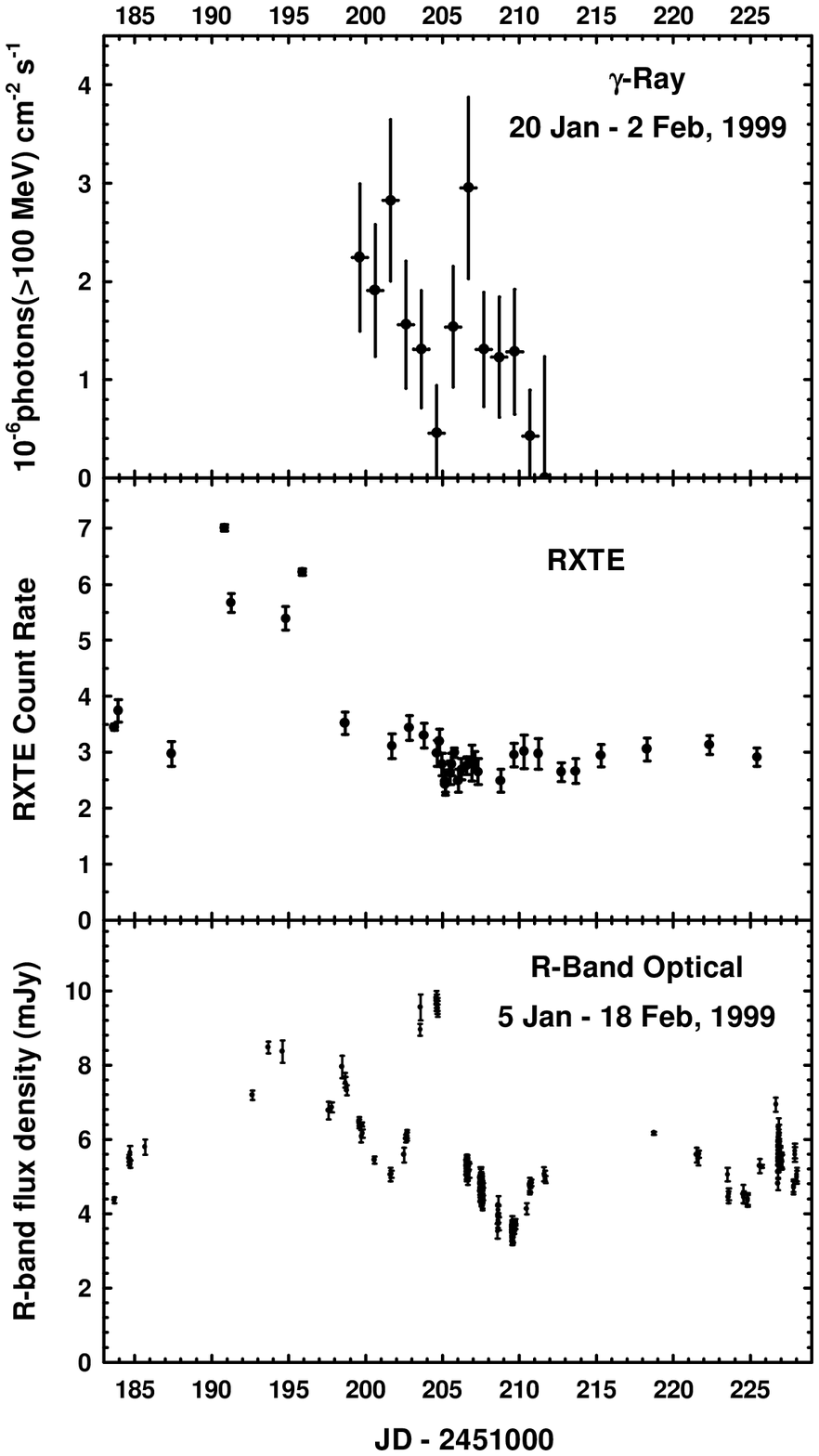}{430pt}{0}{75}{75}{-250pt}{-60pt}
\caption{3C~279 light curves for early 1999 in $\gamma$~rays, X-rays,
and R-band optical}
\end{figure}

\clearpage
\begin{figure}
\figurenum{2}
\epsscale{1.5}
\plotfiddle{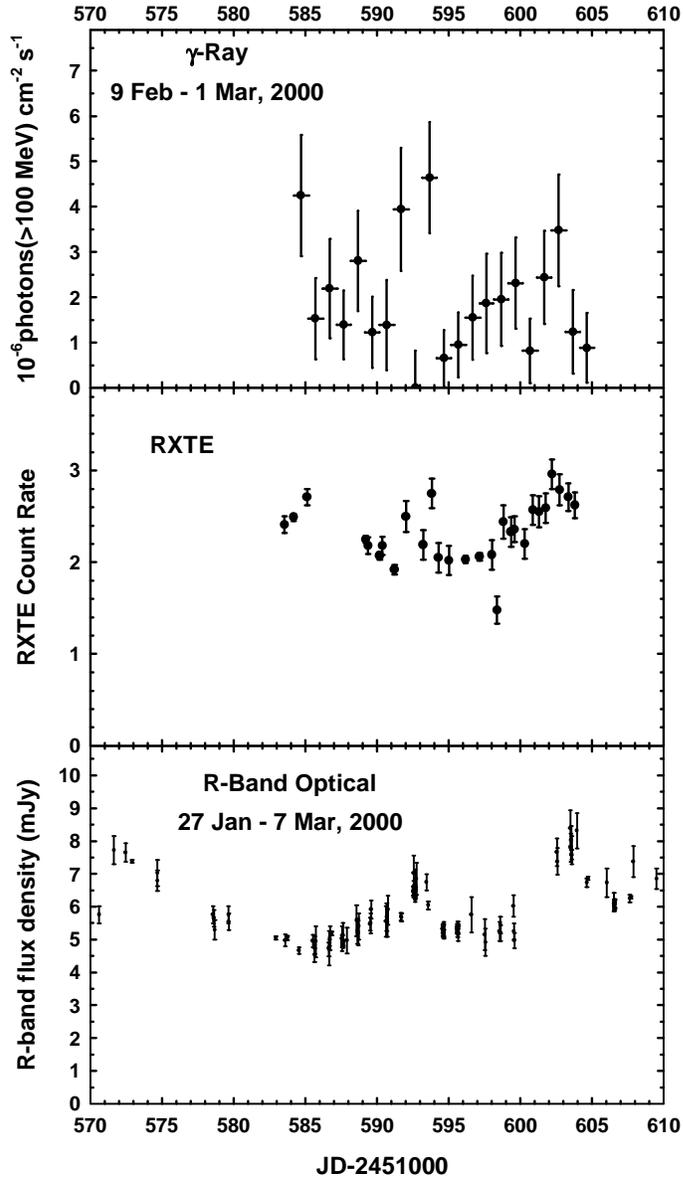}{430pt}{0}{75}{75}{-250pt}{-60pt}
\caption{3C~279 light curves for early 2000 in $\gamma$~rays, X-rays,
and R-band optical}
\end{figure}

\clearpage
\begin{figure}
\figurenum{3}
\epsscale{1.5}
\plotfiddle{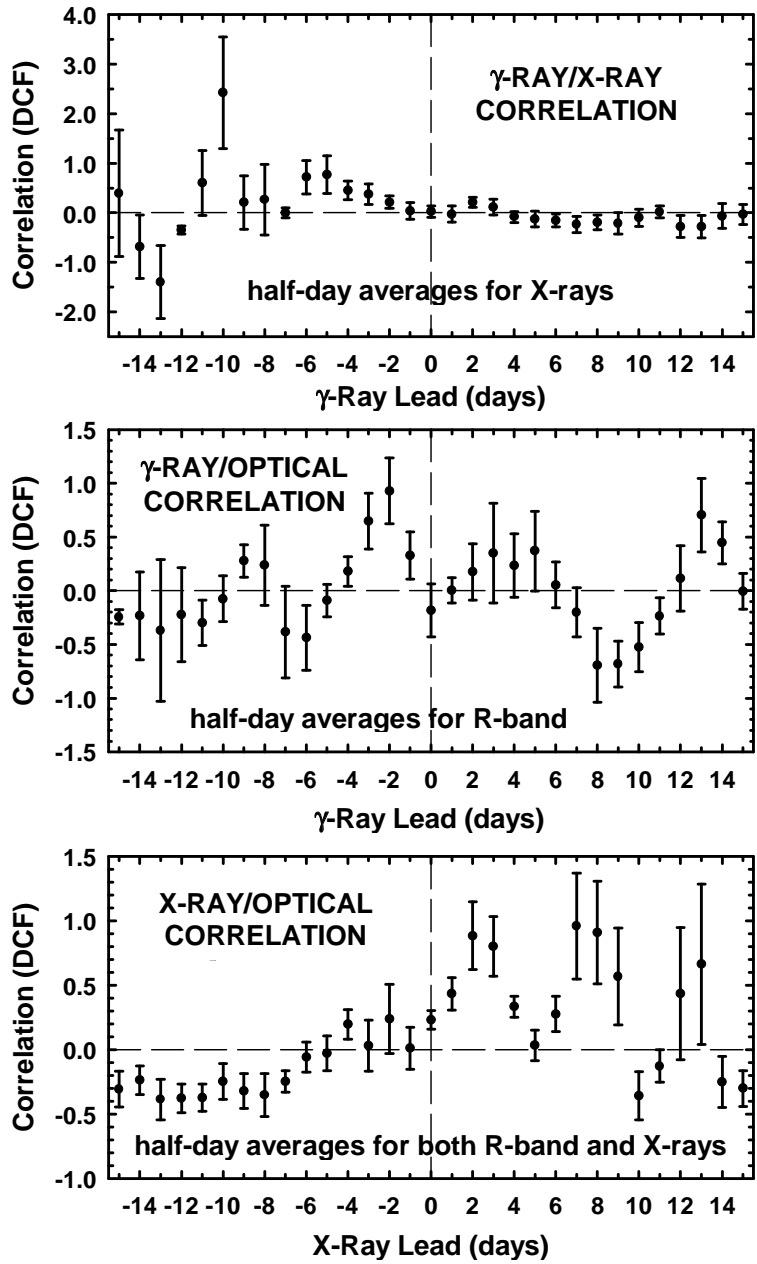}{430pt}{0}{75}{75}{-250pt}{-50pt}
\caption{3C~279 correlation functions (DCF) for 1999}
\end{figure}

\clearpage
\begin{figure}
\figurenum{4}
\epsscale{1.5}
\plotfiddle{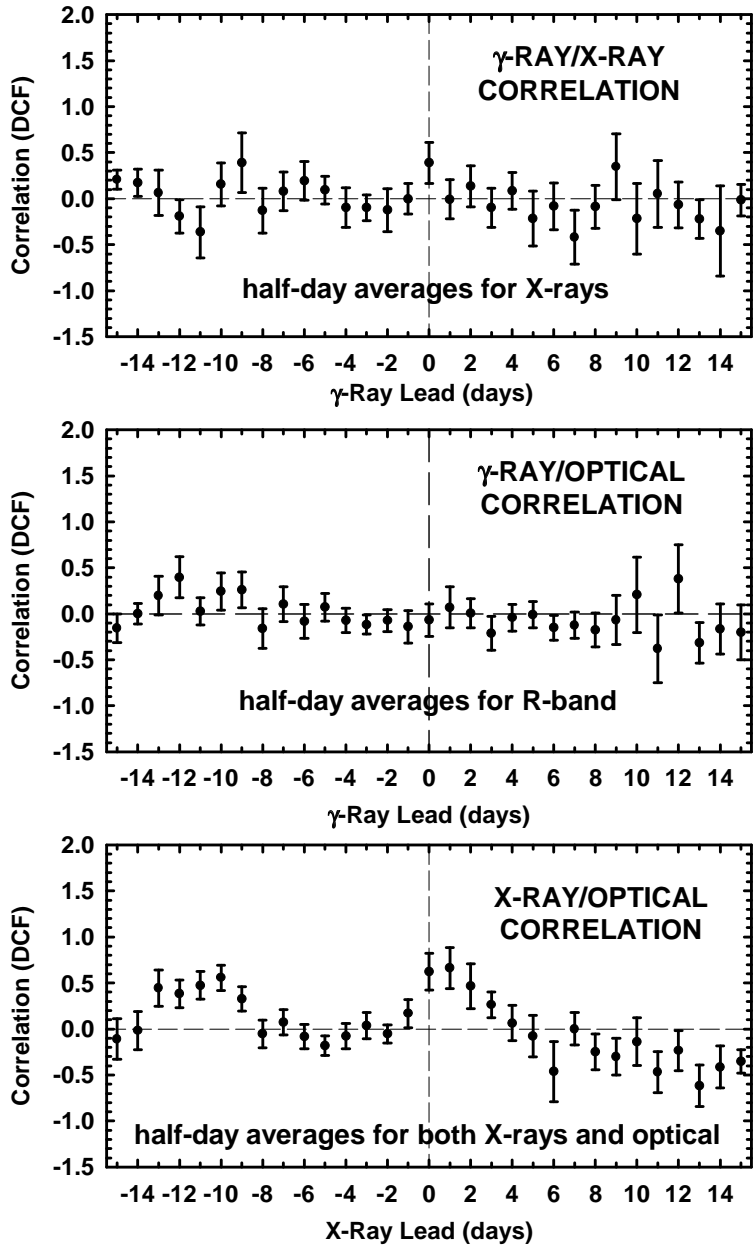}{430pt}{0}{75}{75}{-250pt}{-50pt}
\caption{3C~279 correlation functions (DCF) for 2000}
\end{figure}

\clearpage
\begin{figure}
\figurenum{5}
\epsscale{1.5}
\plotfiddle{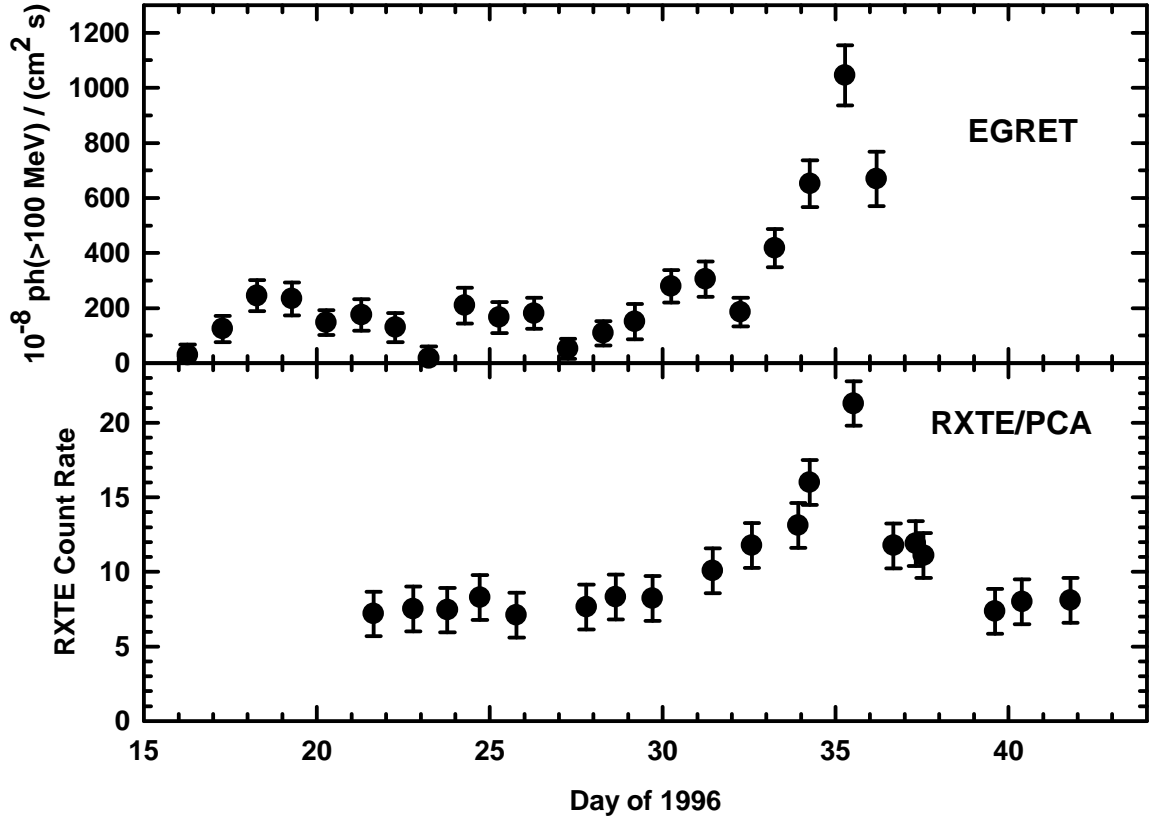}{200pt}{0}{75}{75}{-300pt}{-50pt}
\caption{3C~279 light curves for early 1996 in $\gamma$~rays and
X-rays}
\end{figure}

\clearpage
\begin{figure}
\figurenum{6}
\epsscale{1.5}
\plotfiddle{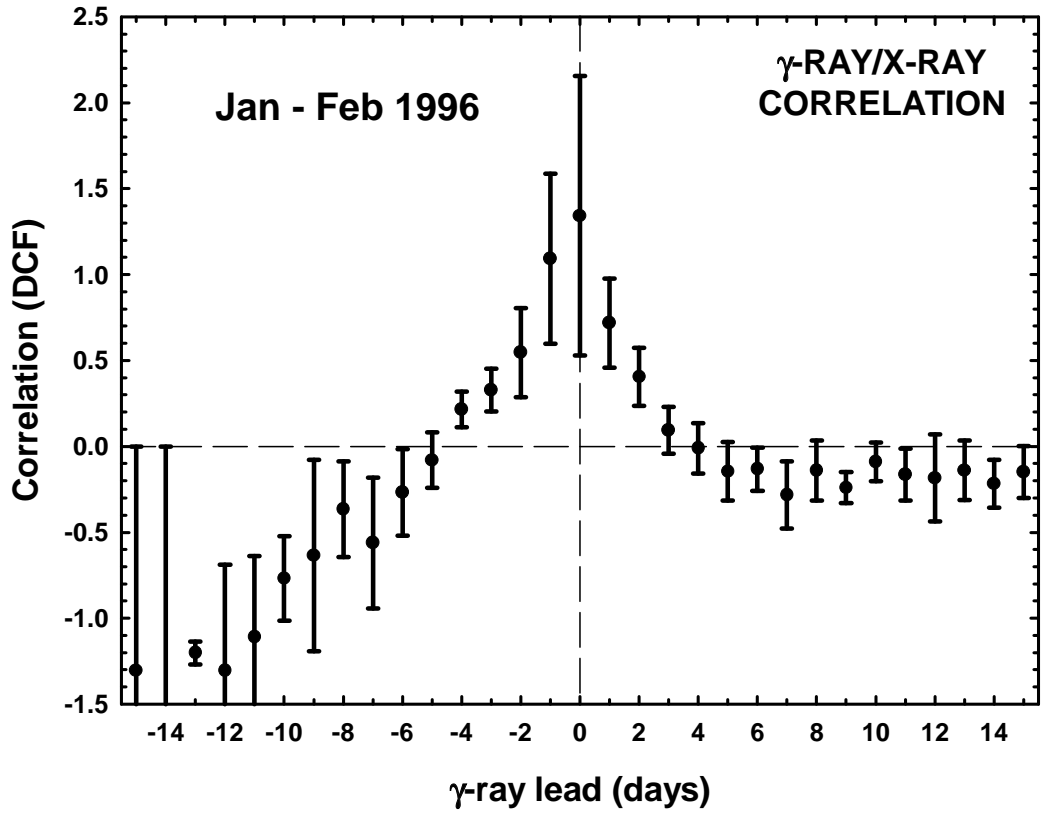}{250pt}{0}{80}{80}{-300pt}{-50pt}
\caption{3C~279 $\gamma$-ray/X-ray correlation function (DCF) for
1996}
\end{figure}

\clearpage
\begin{figure}
\figurenum{7}
\epsscale{1.5}
\plotfiddle{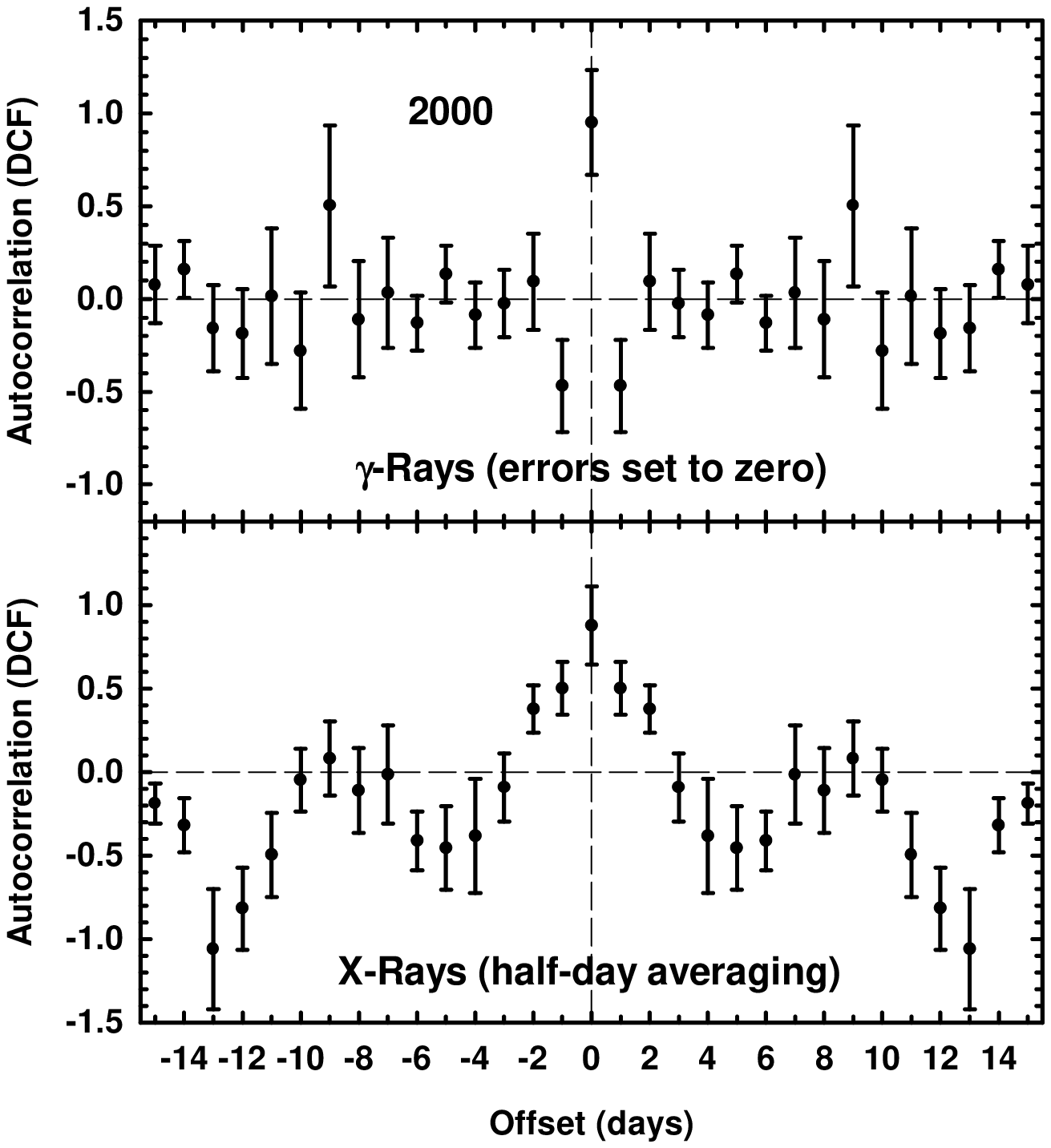}{420pt}{0}{100}{100}{-300pt}{-200pt}
\caption{3C~279 $\gamma$-ray and X-ray autocorrelations for 2000}
\end{figure}


\clearpage

\begin{references}

\reference{} Aharonian, F.A. 2000, New Astron., 5, 377
\reference{} Bertsch, D.L. 2001, poster presentation at ``Gamma 2001''
	conf.; to be published in proceedings (AIP)
\reference{} Blandford, R.D., \& Levinson, A. 1995, \apj, 441, 79
\reference{} Bloom, S.D., \& Marscher, A.P. 1996, \apj, 461, 657
\reference{} Cardelli J.A., Clayton G.C., Mathis J.S., 1989, \apj, 345, 245
\reference{} Dermer, C.D., Schlickeiser, R., \& Mastichiadis, A.
	1992, A\&A, 256, L27
\reference{} Dermer, C.D., \& Schlickeiser, R. 1993, \apj, 416, 458
\reference{} Dermer, C.D., Sturner, S.J., \& Schlickeiser, R. 
	1997, \apjs, 109, 103
\reference{} Edelson, R.A., \& Krolik, J.H. 1988, \apj, 333,646
\reference{} Esposito, J.A., et al. 1999, \apjs, 123, 203
\reference{} Georganopoulos, M., \& Marscher, A.P. 1998, \apj, 506, L11
\reference{} Ghisellini, G., \& Madau, P. 1996, \mnras, 280, 67
\reference{} Hartman, R.C., et al. 2001a, Mem. Soc. Astron. Ital. 
	(in press)
\reference{} Hartman, R.C., et al. 2001b, in Probing the Physics of
        Active Galactic Nuclei by Multiwavelength Monitoring, ed.
	B.M. Peterson, R.S. Polidan, \& R.W. Pogge (San Francisco: 
	Astronomical Society of the Pacific), (in press)
\reference{} Hartman, R.C., et al. 2001c, \apj, 553 (in press)
\reference{} Howell, S.B., \& Jacoby, G.J. 1986, PASP, 98, 802
\reference{} Katajainen, S., Takalo, L.O., \& Sillanp\"a\"a, A., et al. 
	2000, A\&AS, 143, 357
\reference{} Kurtanidze, O.M., \& Nikolashvili, M.G. 1999, Proc.
	of the OJ-94 Annual Meeting 1999, Blazar Monitoring Toward the
	Third Millennium, ed. Raiteri, C.M., Villata, M., \& Takalo, 
	L.O. (Pino Torinese:Osservatorio Astronomico di Torino)
\reference{} Kusunose, M., Takahara, F., \& Li, H. 2000, \apj, 536, 299
\reference{} Lanteri, L., 1999, in OJ-94 Annual Meeting 1999, Blazar 
	Monitoring towards the Third Millennium, ed. Raiteri, C.M., 
	Villata, M., \& Takalo, L.O.(Pino Torinese:Osservatorio
	Astronomico di Torino), 125
\reference{} Li, H., \& Kusunose, M. 2000, \apj, 526, 729
\reference{} Mannheim, K. 1993, A\&A, 269, 67
\reference{} Maraschi, L., Ghisellini, G., \& Celotti, A. 
	1992, \apj, 397, L5
\reference{} Marscher, A.P., \& Gear, W. 1985, \apj, 298, 114
\reference{} Mattox, J.R., et al. 1996, \apj, 461, 396
\reference{} Mattox, J.R., et al. 1999a, in OJ-94 Annual Meeting 1999, 
	Blazar Monitoring towards the Third Millennium, ed. 
	Raiteri, C.M., Villata, M., \& Takalo, L.O.
	(Pino Torinese:Osservatorio Astronomico di Torino), 44 
\reference{} Mattox, J.R., et al. 1999b, PASPC, 189, 95
\reference{} M\"ucke, A., \& Protheroe, R.J. 2001, Astropart. Phys, 15, 121
\reference{} Noble, J. C., Carini, M. T., Miller, H. R., \& Goodrich, B.
        1997, \aj, 113, 1995
\reference{} Protheroe, R.J. 1996a, Adelaide Univ. preprint ADP-AT-96-4
\reference{} Protheroe, R.J. 1996b, Adelaide Univ. preprint ADP-AT-96-7
\reference{} Rachen, J.P. 2000, in ``GeV-TeV Astrophysics: Toward a
	Major Atmospheric Cherenkov Telescope V'', eds. B.D. Dingus 
	et al., AIP Conf. Proc., Vol 515, (AIP:Snowbird),41
\reference{} Raiteri, C.M., Villata, M., Lanteri, L., Cavallone, M., 
	Sobrito, G. 1998, A\&AS, 130, 495
\reference{} Rieke, G.H., \& Lebofsky, M.J. 1985, \apj, 288, 618
\reference{} Sikora, M., Begelman, M.C., \& Rees, M.J. 
	1994, \apj, 421, 153
\reference{} Sikora, M., Blazejowski, M., Begelman, M. C., \& 
	Moderski, R., 2001, \apj (in press)
\reference{} Smith, P. S.  1985, \aj, 90, 1184
\reference{} Takahashi, T., et al. 1996, \apj, 470, L89
\reference{} Thompson, D.J., et al. 1993, \apjs, 86, 629
\reference{} Tosti, G., Pascolini, S., \& Fiorucci, M. 1996,
	PASP, 108, 706
\reference{} Vercellone, S., et al. 1999, in OJ-94 Annual Meeting 1999, 
	Blazar Monitoring towards the Third Millennium, ed. 
	Raiteri, C.M., Villata, M., \& Takalo, L.O.(
	Pino Torinese:Osservatorio Astronomico di Torino), 138
\reference{} Villata, M., et al., 2000, A\&A 363, 108
\reference{} Wehrle, A., et al. 1998, \apj, 497, 178

\end{references}
\end{document}